\documentclass[%
 reprint,
 amsmath,amssymb,
 aps,
 prd
]{revtex4-2}

\usepackage{graphicx}% Include figure files
\usepackage{dcolumn}% Align table columns on decimal point
\usepackage{bm}% bold math
\usepackage{float}
\usepackage{xfrac}
\usepackage{color}
\usepackage[normalem]{ulem}

\begin{document}

\newcommand{\Msun}{M_\odot}
\newcommand{\fderiv}[2]{\frac{\mathrm{d}#1}{\mathrm{d}#2}}
\newcommand{\bcrit}{b_{{\text{crit}}}}

\newcommand{\evan}[1]{{\textcolor{red}{#1}}}
\newcommand{\jeff}[1]{{\textcolor{magenta}{#1}}}

\preprint{APS/123-QED}

\title{Non-Radial Neutrino Emission\\ upon Black Hole Formation in Core Collapse Supernovae}% Force line breaks with \\

\author{Jia-Shian Wang}
\author{Jeff Tseng} % [0000-0003-1731-5853]
\affiliation{%
Department of Physics, Oxford University, Oxford OX1 3RH United Kingdom
}%
\author{Samuel Gullin} 
\author{Evan P. O'Connor}
\affiliation{The Oskar Klein Centre, Department of Astronomy, Stockholm University, AlbaNova, SE-106 91 Stockholm, Sweden}
 % [0000-0002-8228-796X]

\date{\today}

\begin{abstract}
Black hole formation in a core-collapse supernova is expected to lead to a
distinctive, abrupt drop in
neutrino luminosity due to the engulfment of the main neutrino-producing regions
as well as the strong gravitational redshift of those remaining neutrinos which do escape.
Previous analyses of the shape of the cut-off have focused on specific trajectories
or simplified models of bulk neutrino transport.
In this article, we integrate over simple null geodesics to investigate
potential effects on the cut-off profile of including all neutrino emission angles from a
collapsing surface in the Schwarzschild metric, and from a contracting equatorial mass ring
in the Kerr metric.
We find that the non-radial geodesics contribute to a softening of the cut-off
in both cases.  In addition, extreme rotation introduces significant changes to the
shape of the tail which may be observable in future neutrino detectors, or combinations
of detectors.

\end{abstract}

\maketitle

\section{Introduction}

The conventional picture of a core-collapse supernova (CCSN) begins with a
large stellar progenitor, with a zero-age main sequence (ZAMS) mass greater
than approximately 10 solar masses $M_\odot$, collapsing to a neutron star.
In the process, the vast majority of the energy is released in a burst of neutrinos
which escapes the conflagration well before any electromagnetic radiation.
It is expected that neutrino and dark matter detectors will record an abnormally
high rate of neutrino events hours before the supernova becomes visible in the sky;
this lead time is the motivation behind the Supernova Neutrino Early Warning System
(SNEWS)~\cite{Kharusi:2020ovw}, which should play a key role in observing the next
galactic CCSN through its neutrino, gravitational, and electromagnetic ``messengers''.

For some stellar progenitors (though the question of which ones remains debated \cite{Burrows:2020qrp}),
the outcome may be very
different, as the proto-neutron star (PNS) may itself
collapse into a black hole.  In such cases, the neutrino (and gravitational wave)
signal may not be followed by a traditional, visible electromagnetic signal.
One distinctive indication of such an end result is an abrupt cut-off in the neutrino
luminosity.  This cut-off is the result of the black hole engulfing the
neutrino-producing regions of the PNS, and the gravitational redshift of those
which just manage to escape.

The shape of the cut-off was estimated first based on photons emitted from a
free falling, non-rotating mass shell.  A conventional treatment of radial
trajectories gives an exponential decline with a time constant of $4M$, where $M$
is the mass of the black hole~\cite{Hartle2003}.  However, late-time behavior
is expected to be dominated by neutrinos trapped for an extended period around
unstable circular orbits near a critical radius $3M$, resulting in a time constant
of $3\sqrt{3}M$~\cite{podurets:64,AmesThorne:1968}.
In a CCSN, of course, the photons will be absorbed well before they escape.
Instead, the cut-off would only be evident in the neutrinos
escaping from near the growing black hole~\cite{Beacom:2000ng,Beacom:2001}.

Because of uncertainties in the structure of the PNS, it is unclear how much of the
PNS is involved in the creation of the black hole.  It is likely, however, that the
event horizon is formed initially below the surface of the PNS, leaving the most fertile
neutrino production layers outside.  If this is the case, those outside layers will
still be emitting neutrinos in all directions, rather than radially, before they
themselves fall behind the event horizon.  Indeed, it was suggested
in~\cite{Baumgarte:1996iu} that the radial estimate is expected to be an underestimate,
and that a full treatment would require detailed ray-tracing through a highly
curved spacetime.

Nonetheless,
the neutrino cut-off is a compelling feature of black hole formation within a CCSN
because of its simplicity, distinctiveness, and the fact that it could reflect events
occurring deep within one of the most turbulent and violent phenomena known in the
universe.  It can also add statistical power to triangulating the direction of the
CCSN if observed in several detectors~\cite{Beacom:1998fj,Brdar:2018zds,Sarfati:2021vym}.
Models of this phenomenon are usually the domain of computationally intensive
hydrodynamic simulations which of necessity simplify neutrino transport to varying
degrees, and approximate or incorporate full General Relativity
(see, for example, \cite{OConnor2015,Walk:2019miz}).
In many cases, however, black hole formation heralds the end of validity of such
simulations.

In this article, we investigate effects of non-radial geodesics on the neutrino
cut-off for both non-rotating and rotating black holes.  These idealized
trajectories are clearly a drastic simplification within a complex
domain, but as a toy model can highlight broad features and
be a useful check of more detailed simulations.
Null geodesics from collapsing stars have been investigated in
\cite{AmesThorne:1968,LakeRoeder:1979,Shapiro:1989}
for photons, as well as in \cite{DV:1981,DV:1984} for neutrinos,
though in contexts somewhat different from modern models of a CCSN.

This article is organized as follows:
we consider the appropriateness of these simple trajectories
to the core collapse scenario in Section~\ref{sec:model}.
Section~\ref{sec:schwarz} concerns time delays in
the non-rotating Schwarzschild metric.
The Kerr metric, for a rotating black hole, is considered in Section~\ref{sec:kerr}.
The results are summarized and discussed in Section~\ref{sec:conc}.

\section{Neutrinos in an evolving spacetime}
\label{sec:model}

The main source of neutrinos in a CCSN is its hot, dense core.  In the initial
stages of the core collapse, the core undergoes rapid neutronization, leading to a
burst of $\nu_e$.  Later stages see the thermal production of all neutrino flavors within
the coalescing PNS as it continues to accrete matter
from the collapsing star.  Two scenarios are envisaged which result in the
formation of black hole~\cite{Li:2020ujl}:  in the first,
the accretion compresses the PNS further until enough
matter is compressed into a small enough radius that the geodesics of massless
particles curve back on themselves, {\it i.e.}, an event horizon is formed.
In the second, a part of the PNS undergoes a nuclear phase transition,
leading again to higher densities which form an event horizon.

In either case, it is worth noting that the formation of the event horizon,
and therefore of the black hole itself, is in a sense a global observable
irrelevant to a neutrino travelling nearby.  
Assuming that the neutrino can pass through the environment
(a transition which is expected to occur at some point during PNS cooling),
the neutrino is only
sensitive to the local curvature, which is determined by how mass is distributed
around it.  As mass is redistributed around it, generally ``inwards'' in a global
sense, its geodesic bends accordingly.
The only indication that a black hole has formed is
that some (but by no means all) paths bend back on themselves so that they cannot
escape.

Moreover, the timescale on which geodesics bend further is not altogether sudden.  It is
expected that mass accretion onto the PNS will be on the order of $1\Msun$/s 
by the time a black hole forms~\cite{Walk:2019miz}.  If we take this also as an
estimate of how quickly the black hole grows, then its 
outermost circular orbit radius grows at around $3\Msun$/s
(using units in which $G=c=1$), or perhaps more intuitively, on the order of
kilometers per second.  On the other hand, the speed of the neutrinos is
nearly the speed of light, more than $10^5$ greater.  From this perspective,
the slow increase in the size of the black hole should not significantly affect
the neutrino path.

\section{Time delays in the Schwarzschild geometry}
\label{sec:schwarz}

We first consider neutrinos being emitted (or undergoing their last scatter
before escaping) from a shell of matter free-falling radially inward
toward a non-rotating black hole.  We start from
the Schwarzschild metric in the standard coordinates $(t,r,\theta,\phi)$,
\begin{eqnarray}
  \mathrm{d}s^2 & \, = \, &-\left(1 - \frac{2M}{r}\right) \mathrm{d}t^2 +
  \left(1 - \frac{2M}{r}\right)^{-1} \mathrm{d}r^2\nonumber \\ 
  & & + \, r^2\mathrm{d}\theta^2 \, + \, r^2\mathrm{sin}^2\theta \mathrm{d}\phi^2,
\end{eqnarray}
where $M$ is the mass of the black hole.
The geodesic solutions are well known, but we summarize them here.

Since the metric is spherically symmetric, the geodesics lie in an equatorial plane,
for which we choose $\theta = \tfrac{\pi}{2}$.
The conserved quantities associated with the metric's two Killing vectors are
\begin{eqnarray}
  E & = & -g_{\alpha\beta}\left( \frac{\partial}{\partial t} \right)^\alpha
  \left( \frac{\partial}{\partial \tau} \right)^\beta =
  C(r)\frac{ \mathrm{d}t }{ \mathrm{d}\tau }
  \label{eq:sche} \\
  L & = & g_{\alpha\beta}\left( \frac{\partial}{\partial \phi} \right)^\alpha
  \left( \frac{\partial}{\partial \tau} \right)^\beta =
  r^2 \frac{ \mathrm{d}\phi}{\mathrm{d}\tau}
  \label{eq:schl}
\end{eqnarray}
where $\tau$ is the proper time of the neutrino, and
\begin{equation}
    C(r) = \left( 1 - \frac{2M}{r} \right).
\end{equation}
For massive particles, $E$ and $L$ can be interpreted as energy and
angular momentum per unit mass.

\subsection{Neutrino propagation time}

The emitting matter starts from rest at an initial radius $r_0$, from which it
free-falls.  In this case, $L=0$ and we have
\begin{equation}
    \left(\fderiv{r}{t}\right)^2 = \left(1-\frac{C}{C_0}\right)C^2,
    \label{eq:beta}
\end{equation}
where $C_0\equiv C(r_0)$.  For the paths of the nearly massless neutrinos,
we use null geodesics (in which case $\tau$ is technically the affine parameter
rather than the proper time).
The geodesics are solutions of the equation
\begin{equation}
\left(\frac{\mathrm{d}r}{\mathrm{d}t}\right)^2 =
\left(1-\frac{2M}{r}\right)^2
\left(1-\frac{b^2}{r^2}\left(1-\frac{2M}{r}\right)\right),
\label{eq:drdtnull}
\end{equation}
where $b\equiv L/E$ is the impact parameter as observed by a distant observer.

In our model, neutrinos are emitted isotropically in the frame co-moving with the emitter,
which we will refer to as the ``free-falling'' ($FF$) frame.
The emission direction in
the free-falling frame is directly related to the impact parameter defined above
via the frame of an observer static at the radius at which the emission occurs.
We denote this ``static'' frame as $S$.

In the $S$ frame, the (inward) speed of the emitting surface is
\begin{equation}
    \beta_S = \sqrt{1 - \frac{C}{C_0}}.
\label{eq:beta_static}
\end{equation}
The radial velocity of the neutrino upon emission is
\begin{equation}
\left(\frac{\mathrm{d}r}{\mathrm{d}t}\right)_{S} = \cos \psi_{S} =
\sigma_r \sqrt{1-\frac{b^2}{r^2}\left(1-\frac{2M}{r}\right)},
\label{eq:drdtnull_static}
\end{equation}
where $\psi_S$ is the angle relative to the radial outward direction
as measured in the static frame;
and $\sigma_r$ is $+1$ for an outward trajectory, and $-1$ for an inward
trajectory, again as measured in the static frame.
The radial velocity can be identified simply as $\cos \psi_S$,
since for null geodesics the total velocity is always $c = 1$.
This identification also applies to all other frames.

The corresponding emission angle $\psi_{FF}$ in the $FF$ frame is
then related to $\psi_S$ through velocity addition,
\begin{equation}
\begin{split}
    \left(\frac{\mathrm{d}r}{\mathrm{d}t}\right)_{FF} = \cos \psi_{FF} =
    \frac{\cos \psi_S + \beta_S}{1 + \beta_S \cos \psi_S}.
\label{eq:emangle}
\end{split}
\end{equation}
In this way, a value of $b$ can be calculated given an emission angle $\psi_{FF}$.
From Eqs.~\ref{eq:drdtnull_static} and ~\ref{eq:emangle}, one can clearly
see that radial emission corresponds to $b=0$.

Integrating Eq.~\ref{eq:drdtnull} results in the following expression for the travel
time along the geodesic:
\begin{equation}
T(b,r_\ast;r_E) = \int_{r_\ast}^{r_E} \frac{ r^{5/2} \mathrm{d}r }{
(r - 2M) \sqrt{r^3 - b^2(r - 2M)} },
\label{eq:time}
\end{equation}
where $r_\ast$ is the emission radius, and $r_{E}$ is the distance to the Earth.
This expression can be simplified for the case of radial emission
\begin{eqnarray}
    T(b=0,r_\ast;r_E) &= & \int_{r_\ast}^{r_E}\frac{r\mathrm{d}r}{r-2M}\nonumber \\
    &=& (r_E - r_\ast) + 2M\mathrm{ln}\left( \frac{r_E - 2M}{r_\ast - 2M} \right).
\end{eqnarray}

\begin{figure}
    \centering
    \includegraphics[width=8.4cm]{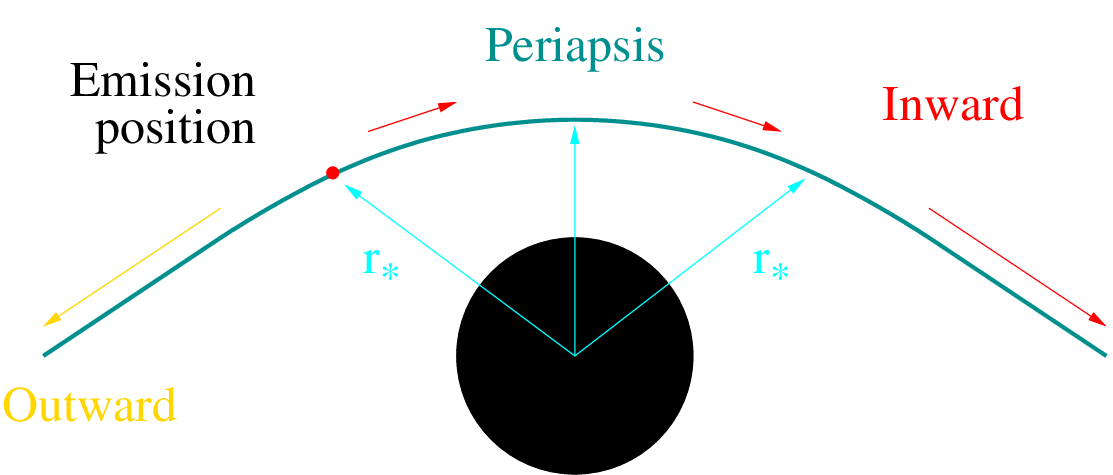}
    \caption{Inward and outward trajectories for neutrinos emitted outside the
    photon sphere.  Outward trajectories escape directly, while inward trajectories
    reach periapsis before joining the outward trajectory.
    }
    \label{fig:peri}
\end{figure}

Fig.~\ref{fig:peri} shows two cases to consider when calculating the
time delay of a neutrino.
For a neutrino emitted in an outward direction (in the static frame $S$),
the time delay relative to the travel time of radial emission from the initial
radius is
\begin{equation}
  \Delta T_+(b, r_\ast) = T(b, r_\ast; r_E) - T(0, r_0; r_E).
  \label{eq:outwardtime}
\end{equation}
If, on the other hand, the neutrino is emitted in an inward direction,
it first acquires a Shapiro-like time delay as it passes the periapsis.  
The travel time is then
\begin{equation}
\Delta T_-(b,r_\ast) = 2T(b,r_p;r_\ast) + \Delta T_+(b,r_\ast),
\label{eq:inwardtime}
\end{equation}
where $r_p$ is the periapsis distance
\begin{equation}
    r_p = \frac{2b}{\sqrt{3}}\cos\left(\frac{1}{3}
    \arccos\left(-\frac{\sqrt{27}M}{b}\right)\right).
\end{equation}
The observation time $t_E$ of a neutrino emitted at time $t_\ast$
from the shell at radius $r_\ast$ can then be written as 
\begin{equation}
    t_E = t_\ast + \Delta T_\pm(b, r_\ast).
    \label{eq:contr}
\end{equation}

\begin{figure}
    \centering
    \includegraphics[width=7cm]{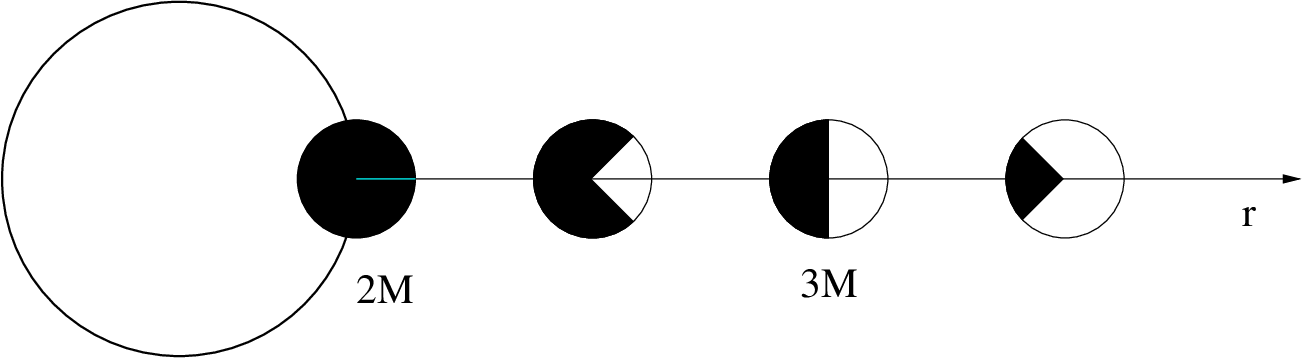}
    \caption{Schematic view of the escape cones (unshaded region) for null geodesics in
    Schwarzschild spacetime.  At $r_\ast = 2M$, only radial geodesics can escape.
    For $2M < r_\ast \leq 3M$, only outward-oriented geodesics are able to escape.
    Beyond $r_\ast = 3M$, some of the inward-oriented geodesics can also escape to infinity.
    }
    \label{fig:escape}
\end{figure}

It is evident from the integrand in Eq.~\ref{eq:time} that not all neutrinos
escape to large distances.
There are two conditions in which the integrand
diverges:  $r=2M$ and $r^3=b^2(r-2M)$.  The first condition gives the event horizon,
while the second condition defines the so-called ``photon sphere'' at $r=3M$,
for which paths with the ``critical impact parameter''
$\bcrit\equiv 3\sqrt{3}M$ follow an unstable circular orbit.
Outside the photon sphere, all outward geodesics, along with inward geodesics
with impact parameters satisfying
\begin{equation}
      \bcrit < b \leq \sqrt{\frac{r^3}{r-2M}},
\end{equation}
are able to reach infinity.  Between the photon sphere and the event horizon,
only outward geodesics with $b<\bcrit$ reach infinity.
Fig.~\ref{fig:escape} shows these ``escape cones'' (the opposite of
Chandrasekhar's ``cones of avoidance''~\cite{Chandrasekhar1983}).

\begin{figure}
    \centering
    \includegraphics[width=\columnwidth]{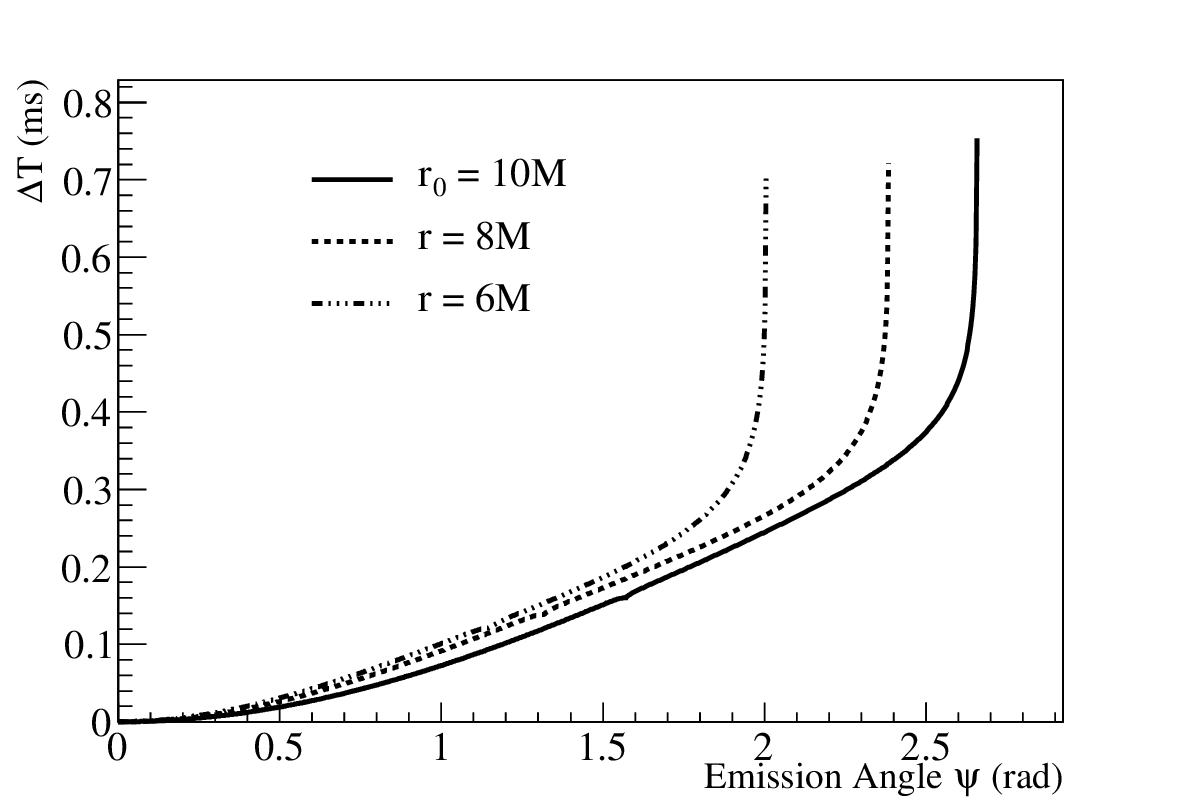}
    \caption{Time delays, relative to that of the outward radial emission
    for a given radius,
    as a function of the emission angle in the $FF$ frame.
    The curves are shown for the initial radius $r_0=10M$ and for subsequent
    radii $8M$ and $6M$, for a Schwarzschild black hole with mass $2.5\Msun$.}
    \label{fig:dt_sch_m25_ff}
\end{figure}

The dependence of the time delay on emission angle is shown in
Fig.~\ref{fig:dt_sch_m25_ff}, for a surface initially at rest
at radius $10M$ and falling towards a black hole of mass $M=2.5\Msun$.
Since $\psi_{FF}=0$ is defined as the outward radial direction, its time delay is zero by
definition.  The delay increases with deviation from the radial direction, and
diverges as the periapsis radius approaches the photon sphere radius.
Moreover, as the emitting shell of matter falls, its velocity increases and soon reaches
speeds which are sizable fractions of $c$; in the given configuration,
the emitters reach a free-fall speed of $0.31c$ at radius $4.3M$,
and pass the photon sphere in 0.55~ms.

\subsection{Luminosity profile}

\begin{figure}
    \centering
    \includegraphics[width=\columnwidth]{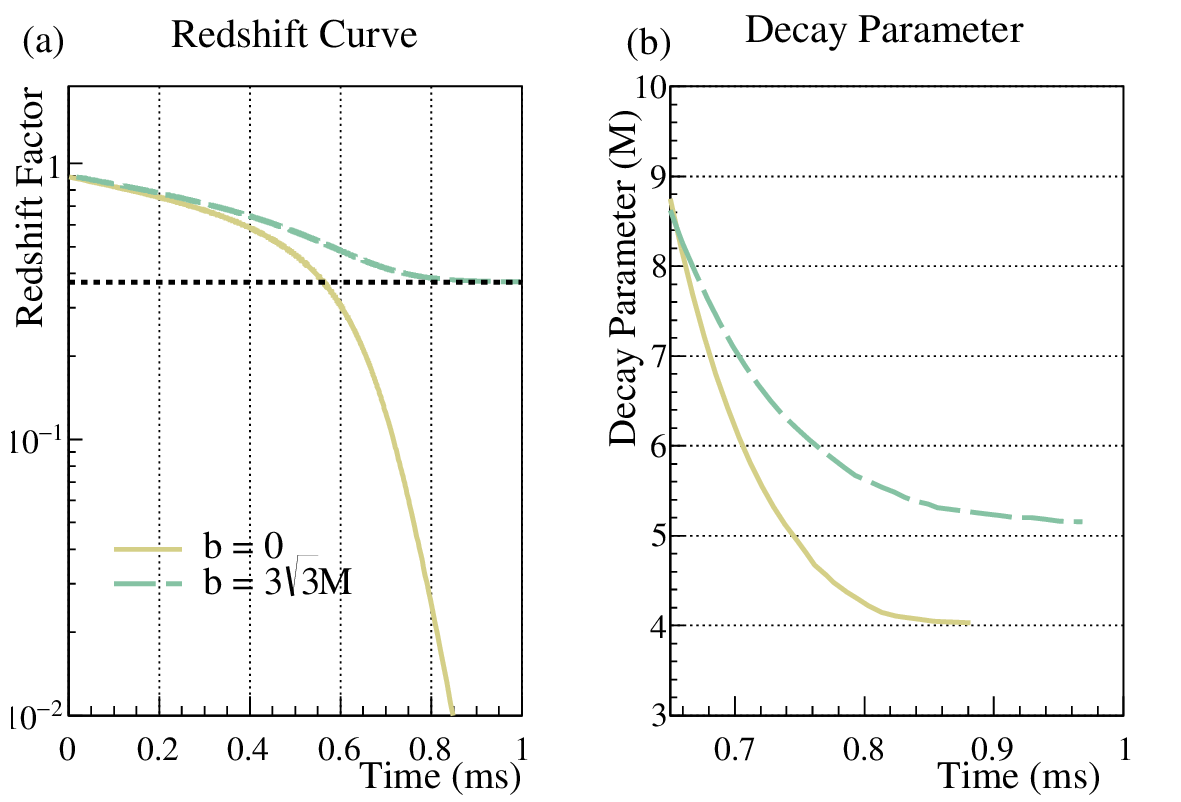}
    \caption{(a) Redshift factor and (b) redshift decay parameter
    for outward radial emissions ($b = 0$, solid/yellow line)
    and critical emissions ($b = \bcrit \equiv 3\sqrt{3}M$, dashed/green line) plotted against
    observation time for a $2.5M_\odot$ Schwarzschild black hole.
    The curves are calculated for a shell falling from $10M$ to $2.1M$
    (for $b=0$) or $3M$ (for $b=\bcrit$).
    The observation time is taken relative to the observation of the
    first neutrino received from the shell when it was at $10M$.
    The dashed black line in (a) is 
    the redshift at the unstable circular orbit at $r = 3M$.
    }
    \label{fig:redshift_radial_m25}
\end{figure}

The cut-off profile of the luminosity as a function of observation time $t_E$
is governed by the number of neutrinos reaching the observer as well as
by their redshift~\cite{AmesThorne:1968}
\begin{equation}
    \zeta(b, r_\ast, \sigma_r) = \frac{\nu_E}{\nu_\ast} =
    \sqrt{C} \times \frac{ \sqrt{1 - \beta_S^2} }{ 1 + \beta_S \cos \psi_S},
    \label{eq:red_rad}
\end{equation}
where $\nu_\ast$ and $\nu_E$ are the neutrino energies at emission and
upon observation at the Earth.
The first factor of Eq.~\ref{eq:red_rad} is the gravitational redshift,
assuming $r_E\gg r_\ast$, and the second factor the Doppler shift.
Fig.~\ref{fig:redshift_radial_m25}(a) shows the redshift factor for outward
radial emissions ($b = 0$) and ``critical emissions'' (those with $b = \bcrit$)
as a function of their observation times,
relative to the arrival of the first neutrino from the shell's initial radius.
The profiles take the form of exponentials with a possible offset,
\begin{equation}
    Ae^{-t/\tau(t)}+B.
\end{equation}
We refer to $\tau(t)$ as the ``decay parameter''
at time $t$, or as the ``(decay) time constant'' when it is indeed constant.
The decay parameter is shown in Fig.~\ref{fig:redshift_radial_m25}(b).
At late times, the radial redshift curve approaches a decay parameter
value of $4M$, consistent with the standard result~\cite{Hartle2003}.
The redshift curve for critical emissions, on the other hand,
flattens out as the shell approaches the circular orbit at radius $3M$;
this feature is analogous to the ``photon cloud'' of \cite{AmesThorne:1968}.
If we subtract off the asymptotic offset, we are left with an
exponential-like attenuation with a decay parameter which approaches
$3\sqrt{3}M$, which is the result of
\cite{podurets:64,AmesThorne:1968}.  Inside radius $3M$, the critical
emissions fall outside the escape cone and do not reach a distant observer.

The energy contribution at observation time $t_E$
of neutrinos emitted at other angles from radius $r$ and time $t$,
with impact parameter $b$, can be written as
\begin{equation}
    \mathrm{d}\epsilon(t_E) = 
    \, \zeta(b, r, \sigma_r) \times \frac{L_0 \cdot \mathrm{d}t}{4\pi r^2}
    \times r^2 \mathrm{d}\Omega,
\label{eq:ws}
\end{equation}
where $\mathrm{d}\Omega$ is the solid angle element of the emitting shell,
and $L_0$ is the total luminosity of the surface.  It is assumed that the
total luminosity is constant throughout the collapse.

\begin{figure}
    \centering
    \includegraphics[width=\columnwidth]{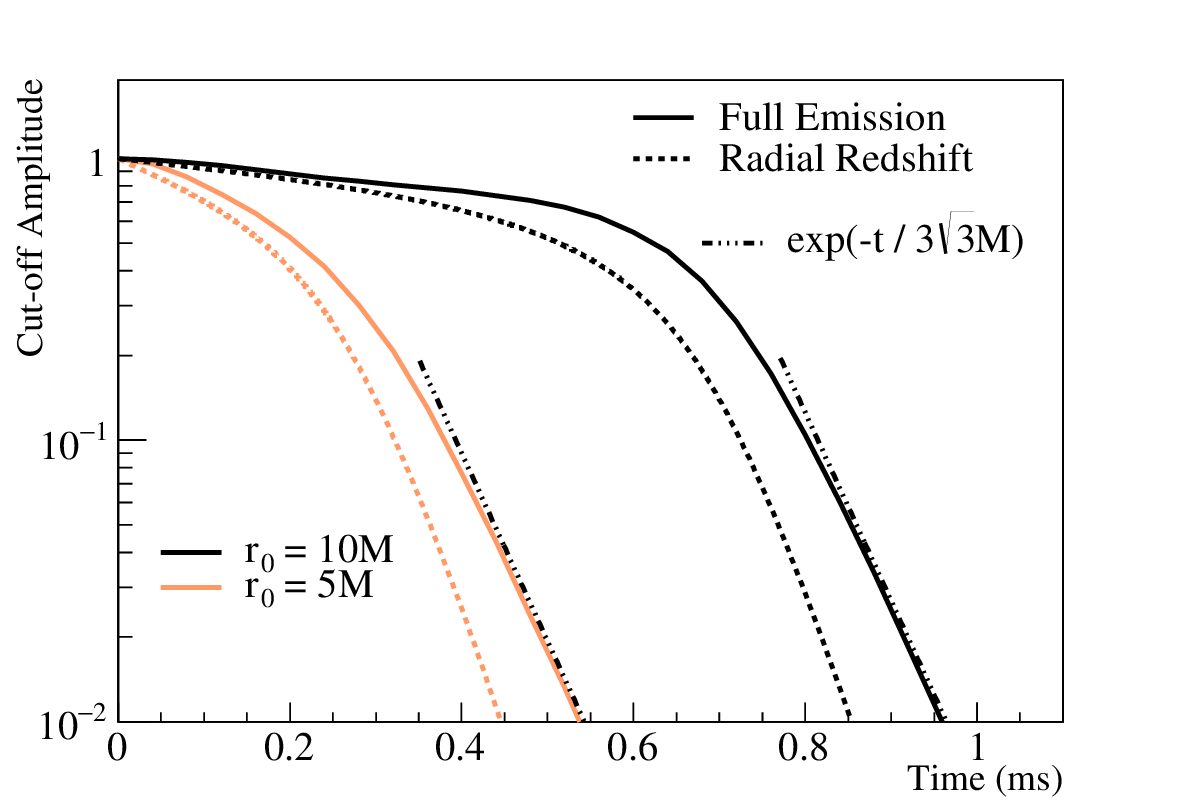}
    \caption{Comparison of cut-off profiles for non-radial (solid) and radial
    (dotted) emissions for initial radii $r_0=10M$ (black) and $5M$ (orange).
    The profiles are normalized to 1 at $t_E=0$.  The late time behaviours
    approach falling exponentials with time constant $3\sqrt{3}M$,
    which are indicated with black dotted-dashed lines.
    }
    \label{fig:m25_mc}
\end{figure}

\begin{figure}
    \centering
    \includegraphics[width=\columnwidth]{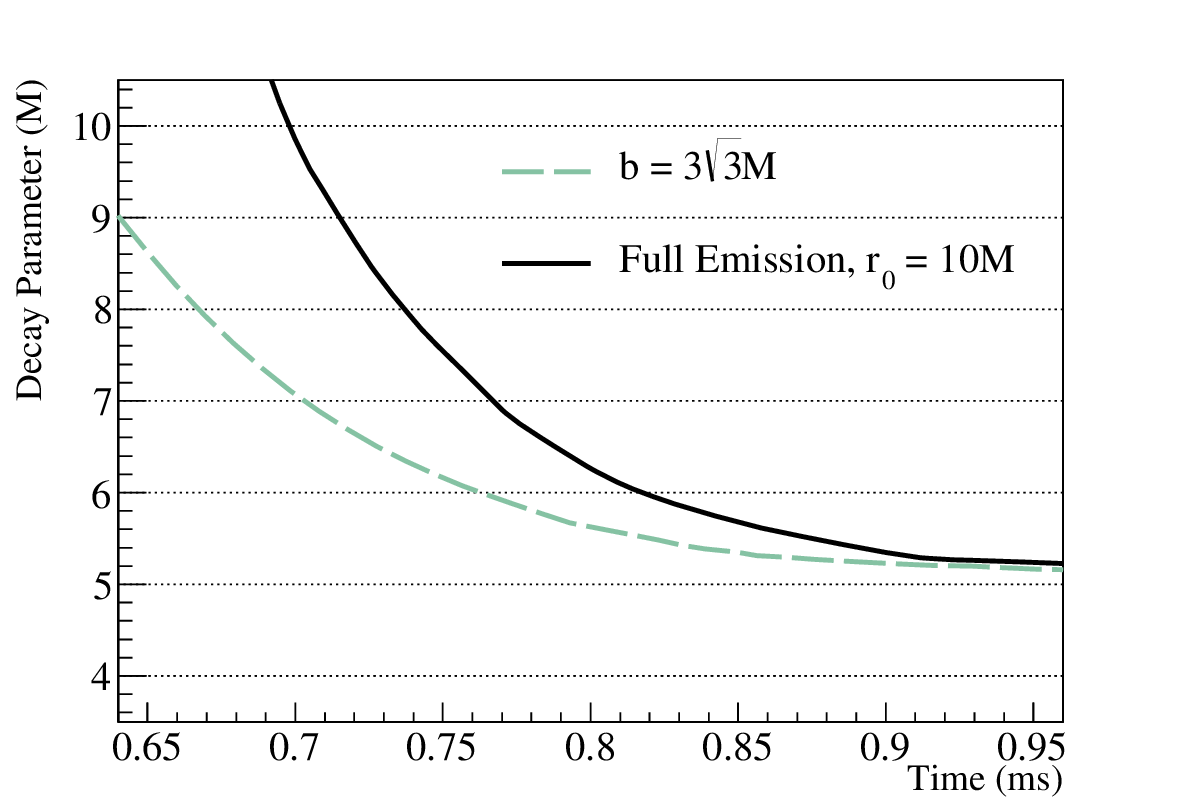}
    \caption{The decay parameter for critical emissions ($b = \bcrit$) (dashed/green)
    and full emissions (solid/black), with initial radius $r_0 = 10M$.  }
    \label{fig:m25_sch_decay}
\end{figure}

We evaluate the luminosity as a function of observed time using a simple
ray-tracing Monte Carlo model.  At each step in coordinate time $t$,
we simulate a fixed number of isotropic neutrino emissions from the collapsing shell.
The emission angle is used to calculate the impact parameter $b$, the corresponding
observation time $t_E$ (if finite), and the redshift factor $\zeta$.
Moreover, we simplify the simulation further in light of its spherical symmetry
by simulating emissions from only one point on the shell, and counting neutrinos
at radius $r_E$, regardless of where the neutrino intersects the outer sphere.
The results are shown in Fig.~\ref{fig:m25_mc} for initial radii
$r_0=10M$ and $5M$:
there is a slow drop in the luminosity for several tenths of ms,
followed by a steepening which rapidly approaches a decay parameter of $3\sqrt{3}M$.
The $3\sqrt{3}M$ decay parameter thus characterizes much of the cut-off
rather than only the very end,
where dominance by critical emissions
near $r=3M$ has long been expected~\cite{Baumgarte:1996iu}.
Moreover, it is evident that consideration of all emission directions
softens the overall cut-off:  Fig.~\ref{fig:m25_sch_decay}, for example,
shows that at all times, the decay parameter of the full emissions curve
exceeds even that of critical emissions.

\begin{figure}
    \centering
    \includegraphics[width=\columnwidth]{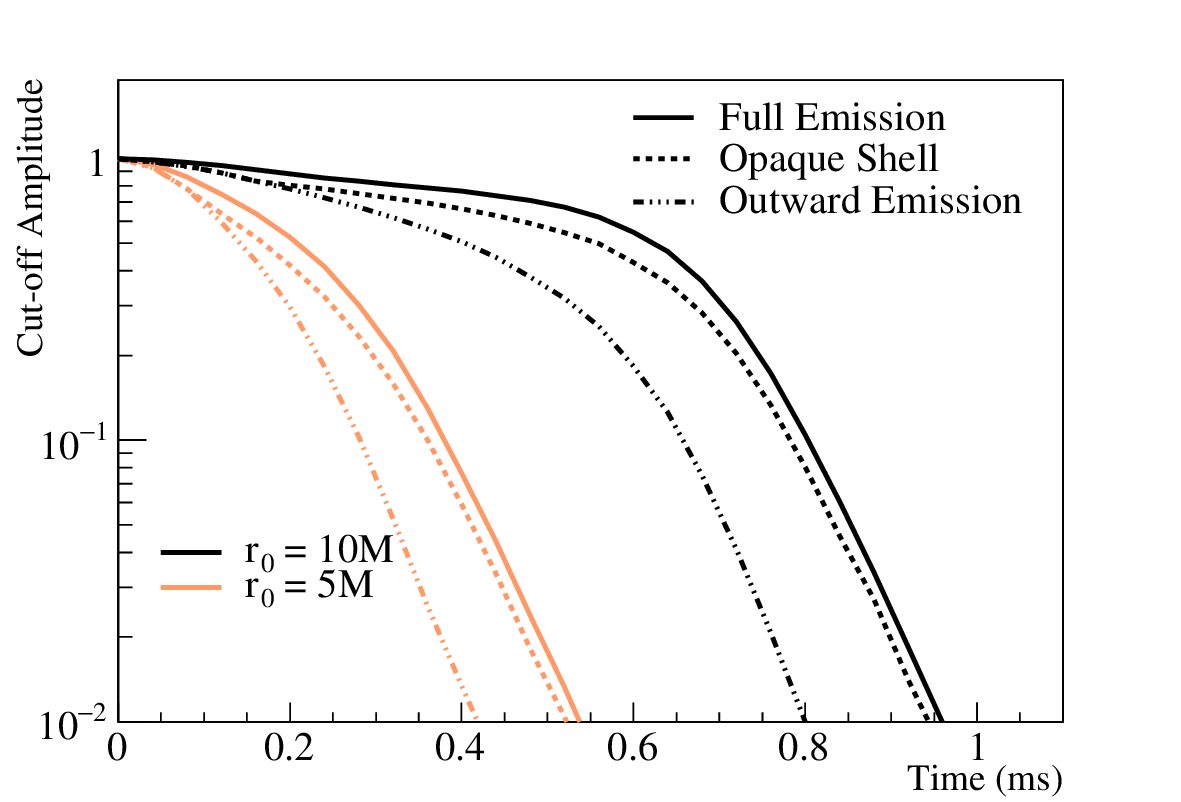}
    \caption{Comparison of three different inward emission scenarios for
    initial radii $r_0=10M$ (black) and $5M$ (orange):
    transparent medium, with emissions allowed in all directions (solid);
    opaque inner medium, allowing only emissions which are outward in the $FF$ frame (dotted);
    and allowing only emissions which are outward in the $S$ frame
    (dashed-dotted).
    }
    \label{fig:m25_opaque}
\end{figure}

It should be noted that Fig.~\ref{fig:m25_mc} shows the result of all the neutrinos
emitted from the collapsing shell and escaping to large distances,
{\it i.e.}, the surrounding medium is transparent to the neutrinos.
If, on the other hand, the inner medium is assumed to be completely opaque to
neutrinos, then only those neutrinos emitted outwards in the free-falling frame $FF$
will escape.  The cut-off profiles of this ``opaque shell'' scenario are shown
in Fig.~\ref{fig:m25_opaque}.  The difference between the opaque shell scenario and
that of allowing only outward emissions in the static ($S$) frame is the
effect of neutrinos which appear in $S$ to be directed inward,
but actually lag behind the collapsing shell and the opaque medium beneath it.
In the end, the opaque shell scenario is a small modification on that of full
emission, and only introduces minor changes to the decay timescale.
%time to fall by $e^{-1}$.

\subsection{Shells not in free fall}
\label{sec:vel}

\begin{figure}
    \centering
    \includegraphics[width=\columnwidth]{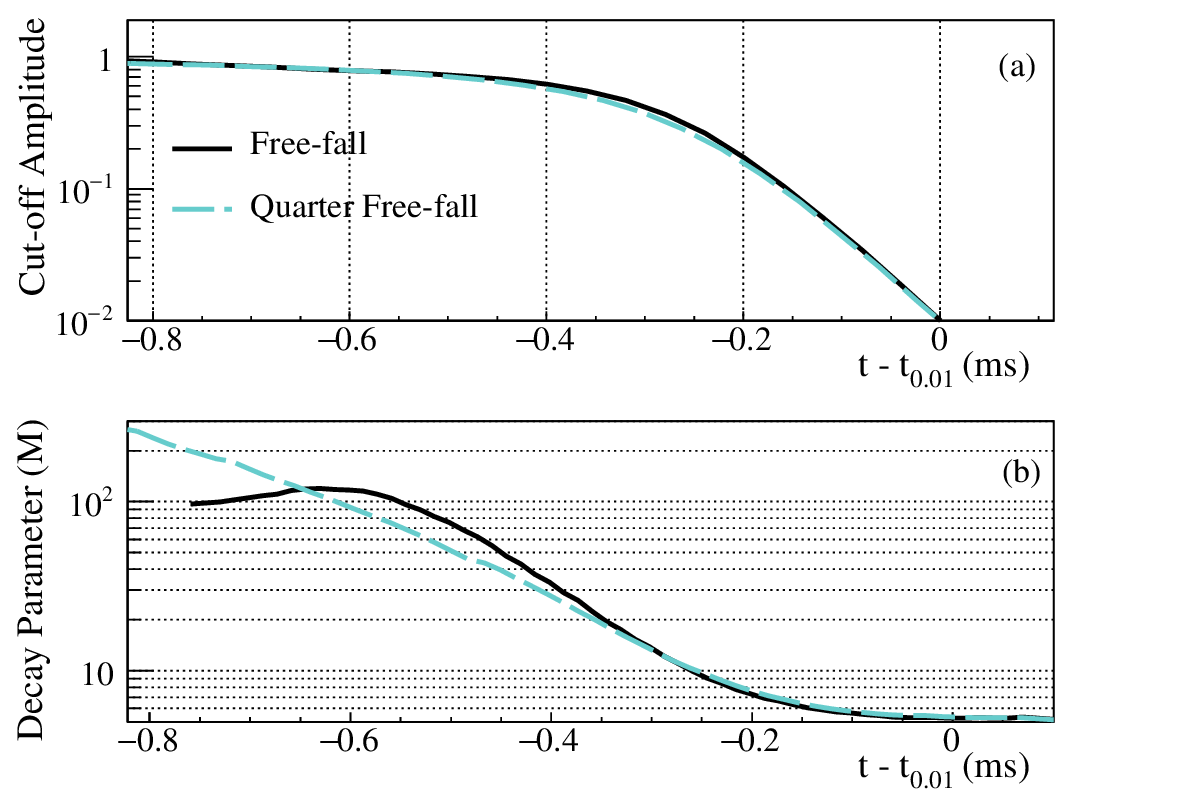}
    \caption{(a) Luminosity and (b) decay parameter profiles for
    free-fall ($f=1$, solid/black) and reduced ($f=1/4$, dashed/blue) shell acceleration.
    The curves are shifted in time to meet at $t=t_{0.01}$, when the luminosity has fallen to 1\%
    of its value at $t=0$.
    }
\label{fig:qaccel}
\end{figure}

The opaque shell is one starting point for introducing more realism into this
toy model.  For instance, we can reduce the shell's proper acceleration by
a constant factor $f$ to mimic residual pressure support.
The proper velocity will be
\begin{equation}
    \frac{\mathrm{d}r}{\mathrm{d}\tau} = \sqrt{f(C_0-C)},
\end{equation}
and the velocity observed in the static frame $S$ will be
\begin{equation}
  \beta_S = \sqrt{ \frac{C_0 - C}{C_0 + \left( f^{-1} - 1\right) C} }.
  %= \sqrt{ \frac{C_0 - C}{C_0 + 3 C} }.
\label{eq:beta_frac}
\end{equation}
The resulting luminosity profile for $f=1/4$, corresponding to a doubling
of the collapse time relative to the free fall case,
is shown in Fig.~\ref{fig:qaccel}.  The slower velocity profile results
in the extension of the slow drop by approximately 0.8ms before the
onset of the rapid decay.  In order to compare the rapid decays, we
shift the times so that the luminosity profiles meet where they have decreased to 1\%
of their starting value; the result is shown in Fig.~\ref{fig:qaccel}.
The quarter free fall decay parameter profile starts from a higher value and
decreases more gradually throughout (the hump near the beginning of the free fall
profile is due to the increasing contribution of inward trajectories).
Once the rapid decay begins, however, it approaches
a decay parameter $3\sqrt{3}M$ in a manner very similar to
the free-fall case.

\begin{figure}
    \centering
    \includegraphics[width=\columnwidth]{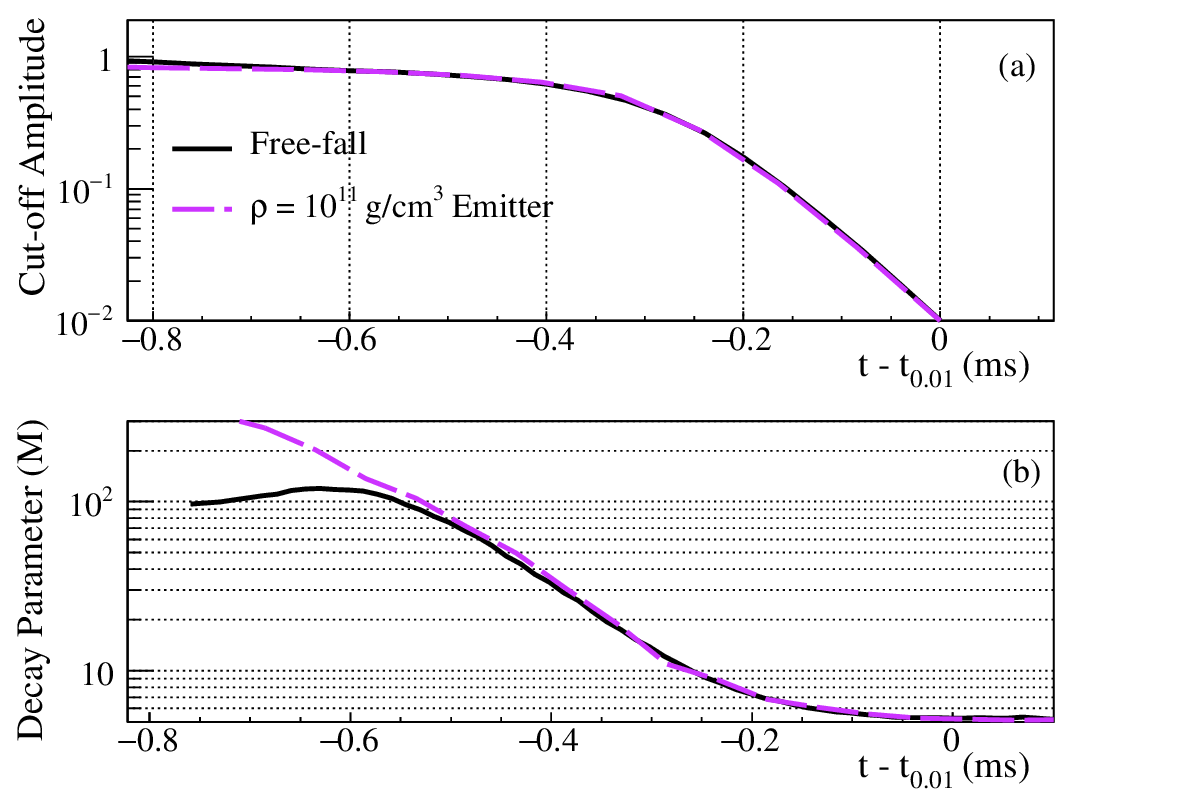}
    \caption{(a) Luminosity and (b) decay parameter profiles for
    free falling and $\rho$-shell emitters.
    The curves are shifted in time to meet at $t=t_{0.01}$, when the luminosity has fallen to 1\%
    of its value at $t=0$.
    }
\label{fig:rho11}
\end{figure}

Another variation of the opaque shell model is to allow the shell radius to
fall at a speed different from that of the emitters themselves.
In this case, the shell is defined by the radius below which the material is dense enough
that the emitted neutrino is expected to scatter or be absorbed.
The emitters, on the other hand, fall through this radius as they emit.
The values of shell radii and emitter velocities as a function of time come
from outside the present toy model, and here we use values from a
GR1D~\cite{O_Connor_2010} general relativistic hydrodynamic simulation, with modern
neutrino transport and interaction rates, of a $40\Msun$ progenitor
model~\cite{woosley:07} (with the Lattimer \& Swesty equation of state with $K_0=220\;{\rm MeV}$),
collapsing to a $2.25\Msun$ black hole~\cite{OConnor2015}.
The shell is defined by density $\rho=10^{11}\;{\rm g}/{\rm cm}^3$,
and simulations have been carried out until the shell has fallen from roughly $7M$ to $3.5M$,
when the simulation ends; the maximum falling speed the shell attains is approximately
$0.35c$.  When the simulation ends, neutrino emission is stopped, though emitted
neutrinos which can escape are propagated to the observer.
The resulting profiles are shown in Fig.~\ref{fig:rho11}.
The luminosity profile remains flatter for longer when compared with the free fall case,
but in the end the decay parameter still approaches
$3\sqrt{3}M$, the result of neutrinos emitted inwards
(though not absorbed by the receding opaque shell) and subsequently trapped near radius $3M$.

\section{Time delays in the Kerr Geometry}
\label{sec:kerr}

In this section, we examine time delays in the Kerr geometry starting from
the formulation of \cite{Igata:2021njn}, which is summarized here.
The metric in Boyer-Lindquist coordinates $(t,r,\theta,\phi)$ is
\begin{eqnarray}
  \mathrm{d}s^2 \, &= &\, -\left(1 - \frac{2Mr}{\Sigma} \right) \mathrm{d}t^2 - \frac{4aMr \mathrm{sin}^2\theta}{\Sigma} \mathrm{d}t\mathrm{d}\phi + \frac{\Sigma}{\Delta}\mathrm{d}r^2\nonumber \\
  & &+ \Sigma \mathrm{d}\theta^2 + \left( r^2 + a^2 + \frac{2Mra^2\mathrm{sin}^2\theta}{\Sigma} \right) \mathrm{sin}^2\theta \mathrm{d}\phi^2,
\end{eqnarray}
where $J$ is the angular momentum of the black hole and
\begin{eqnarray}
  a & \equiv & \frac{J}{M} \\
  \Delta & \equiv & r^2 + a^2 - 2Mr \\
  \Sigma & \equiv & r^2 + a^2\mathrm{cos}^2\theta.
\end{eqnarray}
The Kerr metric is axially symmetric, and, unlike in the Schwarzschild case,
its geodesics do not in general lie in a plane.  Geodesics are
characterized by three constants:  $E$ and $L$ of Eqs.~\ref{eq:sche} and
\ref{eq:schl}, which are shared with the Schwarzschild case; and
$Q$, the Carter constant~\cite{Carter1968}, which can be said to characterize
non-planar motion.
The equations of motion in these coordinates are
\begin{eqnarray}
  \frac{\Sigma}{E} \left( \frac{\mathrm{d}t}{\mathrm{d}\tau} \right) \, &= &\, \frac{1}{\Delta} \left( A - 2Mrab \right) \nonumber \\
  \frac{\Sigma}{E} \left( \frac{\mathrm{d}\phi}{\mathrm{d}\tau} \right) \, &=& \, \frac{b}{\mathrm{sin}^2\theta} + \frac{a}{\Delta} (2Mr - ab)\nonumber \\
  \frac{\Sigma}{E} \left( \frac{\mathrm{d}\theta}{\mathrm{d}\tau} \right) \, &=& \, \sigma_\theta \sqrt{\Theta}\nonumber \\
  \frac{\Sigma}{E} \left( \frac{\mathrm{d}r}{\mathrm{d}\tau} \right) \, &=& \, \sigma_r \sqrt{R}, 
\label{eq:kgeodesics}
\end{eqnarray}
with the useful shorthands 
\begin{eqnarray}
  \Theta &\equiv & q - \mathrm{cos}^2\theta \left( \frac{b^2}{\mathrm{sin}^2\theta} + a^2(\frac{\mu}{E^2} - 1) \right)\nonumber \\
  R &\equiv & \left( (r^2+a^2) - ab \right)^2 - \Delta \left( \frac{\mu}{E^2} r^2 + q + (a - b)^2 \right)\nonumber \\
  A &\equiv &  (r^2 + a^2)^2 - a^2\Delta \sin^2 \theta, 
\end{eqnarray}
where $\mu = 0$ gives null and $\mu = 1$ time-like geodesics,
and $\sigma_r$ and $\sigma_\theta$ indicate the direction of motion
relative to the $r$ and $\theta$ axes.
We define impact parameters $b\equiv L/E$, as before, and
$q\equiv Q/E^2$.  Radial directions coincide with $b=q=0$.  We
also identify co-rotating geodesics with $b>0$, and
counter-rotating with $b<0$.

\subsection{Neutrino propagation time}

As with the non-rotating case, we model a thin shell of emitting matter
falling freely from rest from an initial radius $r_0$.
Instead of the frame of a static observer, we use the locally non-rotating
frame (LNRF)~\cite{Bardeen1972}.  An observer static in the LNRF is known as a
zero angular momentum observer (ZAMO), as its four-velocity gives $L=0$.

The free-falling emitter in the Kerr geometry coincides with the ZAMO at $r=r_0$,
at which location the emitter is initially at rest.
Since, in the LNRF,
\begin{equation}
    \left(\frac{\mathrm{d}t}{\mathrm{d}\tau}\right)_{LNRF} =
    \sqrt{\frac{A}{\Sigma\Delta}},
\end{equation}
one finds the constants of motion
\begin{equation}
  Q = a^2(1 - E^2)\cos^2\theta
\end{equation}
and 
\begin{equation}
  E = \sqrt{ \frac{\Sigma_0 \Delta_0}{A_0} },
\end{equation}
where $\Sigma_0 = \Sigma(r_0)$, $\Delta_0 = \Delta(r_0)$ and $A_0 = A(r_0)$.  
In the coordinates of the distant observer,
\begin{equation}
  \frac{\mathrm{d}r}{\mathrm{d}t} = \frac{\Delta}{\sqrt{A}} \cdot
  \sqrt{1 - \frac{ \Sigma \Delta / A }{ \Sigma_0 \Delta_0 / A_0 }}
\end{equation}
which in the LNRF becomes
\begin{equation}
  \left(\frac{\mathrm{d}r}{\mathrm{d}t}\right)_{LNRF} =
  \sqrt{1 - \frac{ \Sigma \Delta / A }{ \Sigma_0 \Delta_0 / A_0 }}.
  \label{eq:beta_lnrf}
\end{equation}
These correspond to the velocities in Eqs.~\ref{eq:beta} and \ref{eq:beta_static}.
For convenience in this Section, we re-purpose the subscript $S$ to represent the quantities
observed in the LNRF.  The velocity in Eq.~\ref{eq:beta_lnrf}
will therefore be denoted as $\beta_S$.

As before, we relate the emission angles in the free-falling
frame $FF$ to constants of the neutrino's subsequent geodesic.  The angle $\psi_S$
is defined relative to the outward radial direction in the LNRF,
\begin{equation}
  \left(\frac{\mathrm{d}r}{\mathrm{d}t}\right)_{LNRF} =
  \cos \psi_S = \frac{ \sigma_r \sqrt{RA} }{A - 2Mrab}.
\end{equation}
The corresponding angle $\psi_{FF}$ in the $FF$ frame can
then be written in the same form as Eq.~\ref{eq:emangle}.

In the Kerr case, however, geodesics are not necessarily planar, and we define
an out-of-plane angle $\eta$ as the azimuthal angle around the outward radial
direction as the axis, with $\eta=0$ denoting the positive $\theta$ direction.
In this way, $\eta\in[0,\pi)$ indicates a trajectory co-rotating with the
black hole, and $\eta\in[\pi,2\pi)$ counter-rotating.
In the LNRF, this definition gives
\begin{equation}
  \left(r\frac{\mathrm{d}\theta}{\mathrm{d}t}\right)_{LNRF} =
  \sin \psi_S \cos \eta_S = \frac{ \sigma_\theta \sqrt{\Theta \Delta A} }{A - 2Mrab},
\end{equation}
whereas in the $FF$ frame,
\begin{equation}
  \left(r\frac{\mathrm{d}\theta}{\mathrm{d}t}\right)_{FF} =
  \sin \psi_{FF} \cos \eta_{FF} =
  \frac{ \sqrt{1 - \beta_S^2} \sin \psi_S \cos \eta_S }{ 1 + \beta_S \cos \psi_S}.
\end{equation}
Hence, for a given direction $(\psi_{FF},\eta_{FF})$ in the $FF$ frame,
we calculate the quantities
\begin{eqnarray}
  B &\equiv & a^2 - \frac{\Delta}{\sin^2 \theta}\nonumber \\
  \kappa^2 &\equiv & (\upsilon_r)_S^2 + (\upsilon_\theta)_S^2 = \cos^2 \psi_S + \sin^2 \psi_S \cos^2 \eta_S\nonumber \\
  (\upsilon_r)_S &=& \cos \psi_S = \frac{\cos \psi_{FF} - \beta_S}{1 - \beta_S \cos \psi_{FF}}\nonumber \\
  (\upsilon_\theta)_S &=& \sin \psi_S \cos \eta_S\nonumber \\
  &=& \frac{ \sin \psi_{FF} \cos \eta_{FF} (1 + \beta_S \cos \psi_{FF})}{\sqrt{1 - \beta_S^2}},
\end{eqnarray}
from which we calculate the constants of the neutrino path
\begin{equation}
  b = \frac{ A\left(2Mra(1-\kappa^2) - \sigma_b \sqrt{ (1 - \kappa^2)(4M^2r^2a^2 - AB) } \right)}{AB - 4M^2r^2a^2\kappa^2},
\end{equation}
where $\sigma_b \equiv b/|b|$ is the rotating direction relative to the rotation
of the black hole, and
\begin{equation}
  q = \cos^2 \theta \left(\frac{b^2}{\sin^2 \theta} - a^2 \right) + \sin^2\psi_S \cos^2\eta_S \frac{(A - 2Mrab)^2}{\Delta A}.
\end{equation}

We can now calculate the
travel time in a manner similar to Eq.~\ref{eq:time} using the integral
\begin{equation}
  T(b,q,r_\ast; a,r_E) = \int_{r_\ast}^{r_E} \frac{ A - 2Mrab }{\sigma_r \sqrt{R} \Delta}
  \mathrm{d}r.
\label{eq:ktime}
\end{equation}
Since the integrand has an implicit dependence on $\theta$ via $A$,
we propagate both $r$ and $\theta$ along the geodesic by equating integrals of
the last two equations of Eq.~\ref{eq:kgeodesics}:
\begin{equation}
  \int_{r_\ast}^{r} \frac{\mathrm{d}r}{\sigma_r \sqrt{R}}  =
  \int_{\theta_\ast}^{\theta} \frac{\mathrm{d}\theta}{\sigma_\theta \sqrt{\Theta} }. 
\end{equation}
For null geodesics, the $\theta$ integral is evaluated with
\begin{equation}
  \int_{\theta_\ast}^{\theta} \frac{\mathrm{d}\theta}{\sqrt{\Theta} } =
  \int_{\cos \theta_\ast}^{\mathrm{cos}\theta} \frac{ -\mathrm{d\cos\theta}}{
  \sqrt{q + (a^2 - q - b^2)\mathrm{cos}^2\theta - a^2\mathrm{cos}^4\theta} }.
\end{equation}
The time delays are calculated in the same fashion as
Eqs.~\ref{eq:outwardtime} and \ref{eq:inwardtime}, albeit with an extra dependence on $q$.
For the inward case, the periapsis is the largest real root of $R$ at some given $b$ and $q$.

The escape conditions for the Kerr metric are more complicated than
those for the Schwarzschild metric and entail a number of cases which are tabulated
in~\cite{Ogasawara:2020frt} for a full Kerr space and
\cite{Igata:2021njn} for a disc model.

\subsection{Luminosity profile}

Following the same procedure for deriving the gravitational redshift in the
Schwarzschild case, we find the gravitational redshift to be
\begin{equation}
  \frac{\nu_E}{\nu_S} = \sqrt{ \frac{\Sigma \Delta}{A} },
\end{equation}
where $\nu_E$ and $\nu_S$ are the energies observed on Earth and in the
LNRF at the emission position.
The total redshift factor is then
\begin{equation}
  \zeta(b, q, r_\ast, \theta_\ast, \sigma_r) = \frac{\nu_E}{\nu_\ast} =  \sqrt{ \frac{\Sigma \Delta}{A} } \times \frac{ \sqrt{1 - \beta_S^2} }{ 1 + \beta_S \cos \psi_S},
\end{equation}
where $\nu_\ast$ is the neutrino energy in the $FF$ frame of the emitter.

\begin{figure}
    \centering
    \includegraphics[width=\columnwidth]{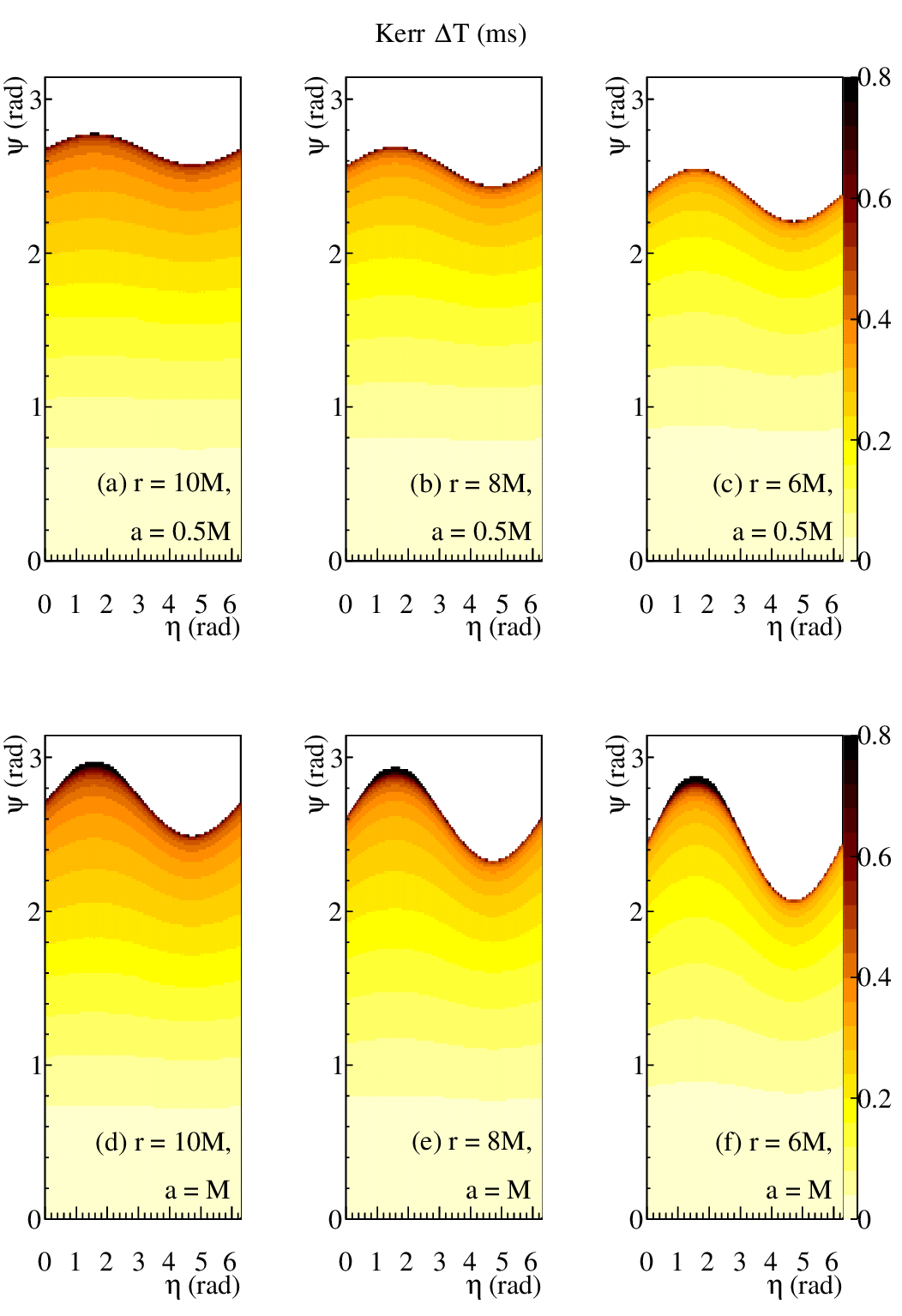}
    \caption{Time delays (color or gray scale, in units of milliseconds),
    relative to that of the radial direction,
    as a function of the emission direction in the $FF$ frame,
    at different emission radii for a collapsing ring around a rotating
    black hole.
    The white region at large $\psi_{FF}$ indicates non-escaping directions;
    in some areas the black region, which always borders on a white region,
    is too narrow to show up in the figure.
    The black hole mass is
    $M=2.5\Msun$, and the ring starts from rest at $r_0=10M$.
    Top row: sub-extremal rotation $a=0.5M$.
    Bottom row:  extremal rotation $a=M$.}
    \label{fig:dt_ker}
\end{figure}

\begin{figure}
\includegraphics[width=\columnwidth]{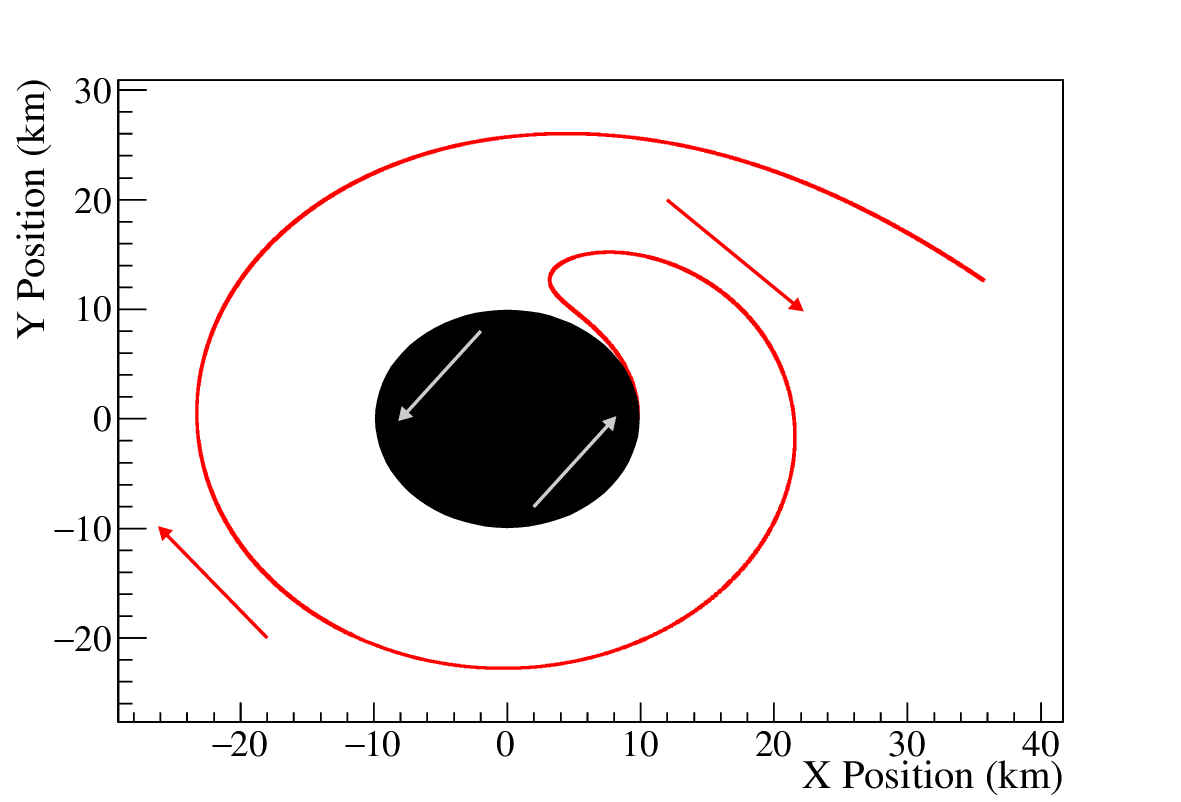}
\caption{An example geodesic for a particle emitted close to the event horizon
of a rotating black hole.  This geodesic is classified as ``counter-rotating''
due to strong local frame dragging near the horizon,
{\it i.e.}, the trajectory is opposite the direction of rotation in its LNRF,
though not in the coordinate system of a distant observer.
}
\label{fig:reverse}
\end{figure}

\begin{figure}
    \centering
    \includegraphics[width=\columnwidth]{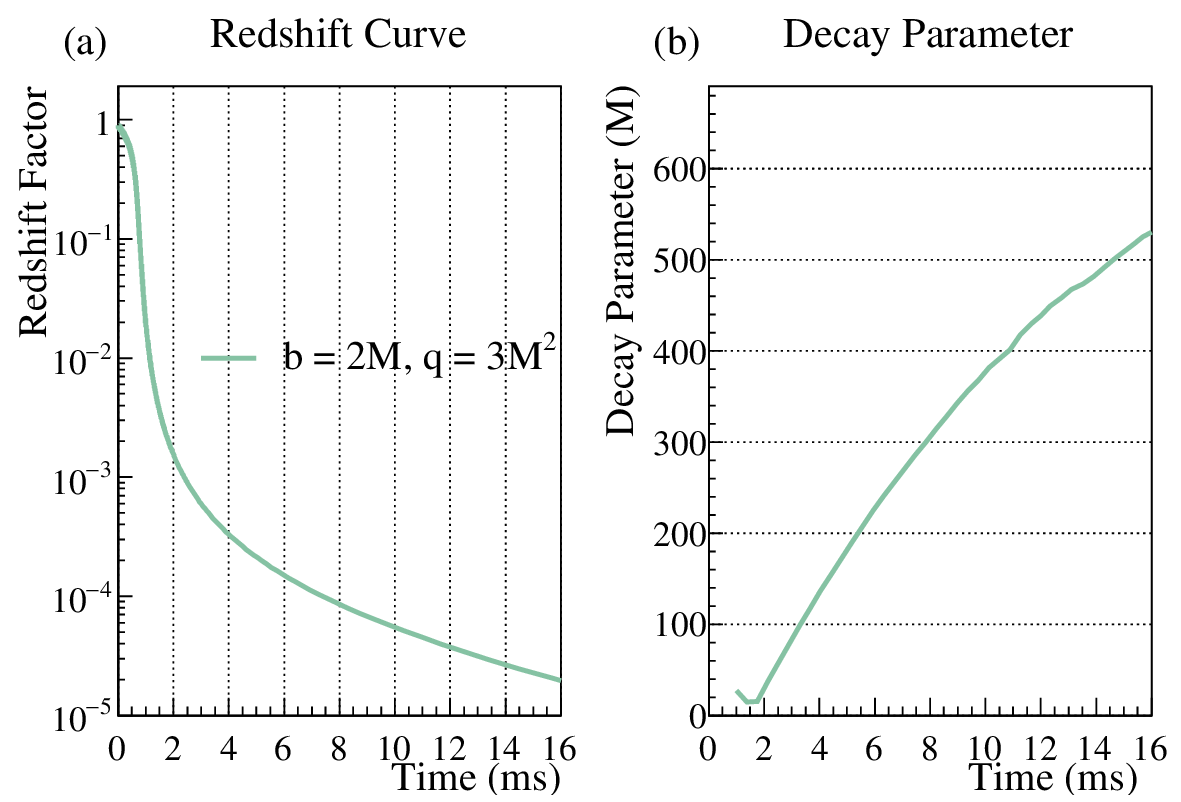}
    \caption{(a) Redshift factor and (b) redshift decay parameter of emissions
    with $b = 2M$ and $q = 3M^2$, for orbits near the 
    horizon at $r=M$, plotted against observation time for a
    $2.5M_\odot$ extremal ($a = M$) Kerr black hole.  
    }
    \label{fig:ker_red_limit}
\end{figure}

We use the same ray-tracing Monte Carlo approach as before.  In order to reduce
the simulation time, however, we only calculate results for emissions from
a contracting ring of matter in the equatorial plane.
Fig.~\ref{fig:dt_ker} shows the time delays
(relative to that of radial emission from the given radius) as
a function of $\psi_{FF}$ and $\eta_{FF}$, for rotation parameters $a=0.5M$ and $a=M$.
The rotation in each case is in the direction of increasing azimuthal angle $\phi$.
Rotation introduces an asymmetry in $\eta_{FF}$:  at a given $\psi_{FF}$, counter-rotating
geodesics, with $\eta_S\in(\pi,2\pi)$, tend to undergo longer delays (note that the
white regions in Fig.~\ref{fig:dt_ker} indicate trajectories with above-critical
impact parameters, {\em i.e.}, diverging escape times).
The delay is especially long for initially counter-rotating geodesics reversing direction before
escaping, as illustrated in one example in Fig.~\ref{fig:reverse}.
A further example of the effect of such geodesics is shown in
Fig.~\ref{fig:ker_red_limit}, which shows the redshift curve and decay parameter,
as a function of observation time, for emissions with $b=2M$ and $q=3M^2$ originating
near the horizon at $r=M$  of an extremal Kerr black hole.
As seen in the decay parameter curve,
such geodesics ``leak'' out very slowly compared to the usual cut-off timescale
of $O(0.1)$~ms in a manner similar to the leakage from the Schwarzschild black hole's
``photon sphere'' at $r=3M$.

\begin{figure}
    \centering
    \includegraphics[width=\columnwidth]{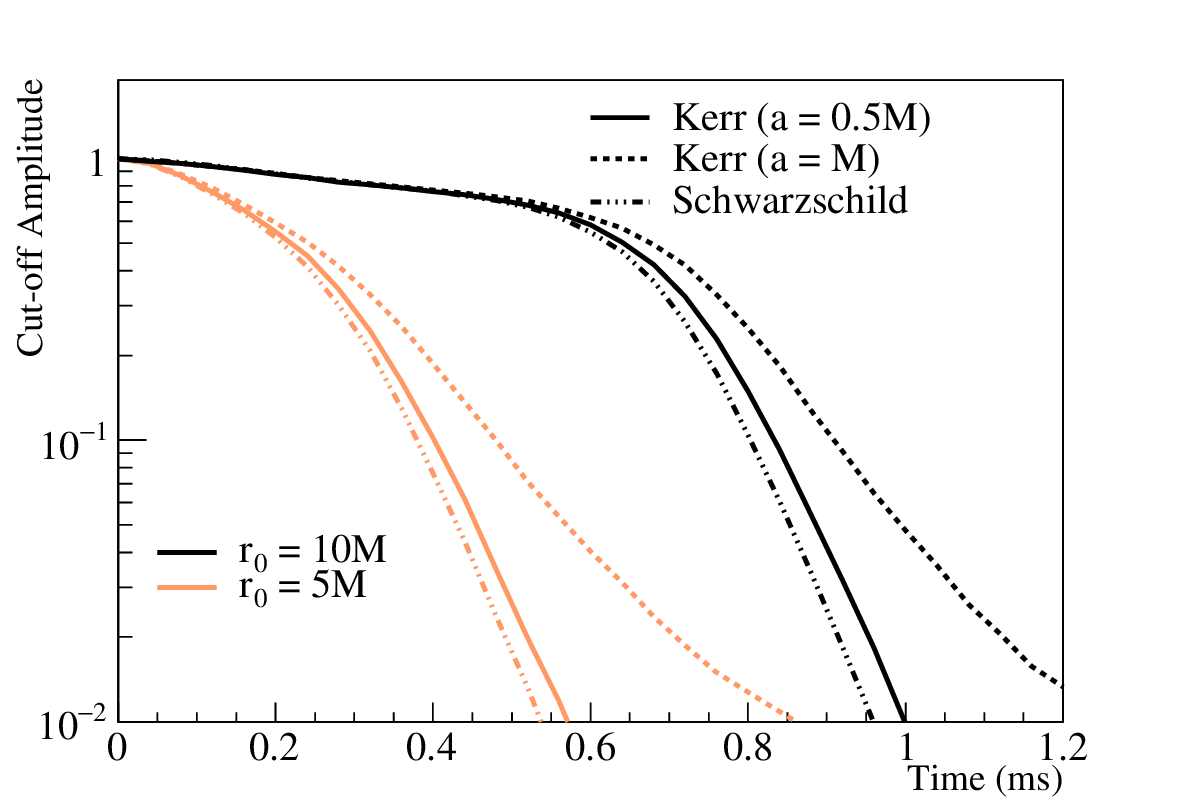}
    \caption{Luminosity profiles of the neutrino cut-off in the Kerr geometry,
    with $a=0.5M$ and $a=M$, and at different initial radii $r_0=5M$ and $10M$.}
    \label{fig:cutoff_kerr_m25}
\end{figure}

\begin{figure}
    \centering
    \includegraphics[width=\columnwidth]{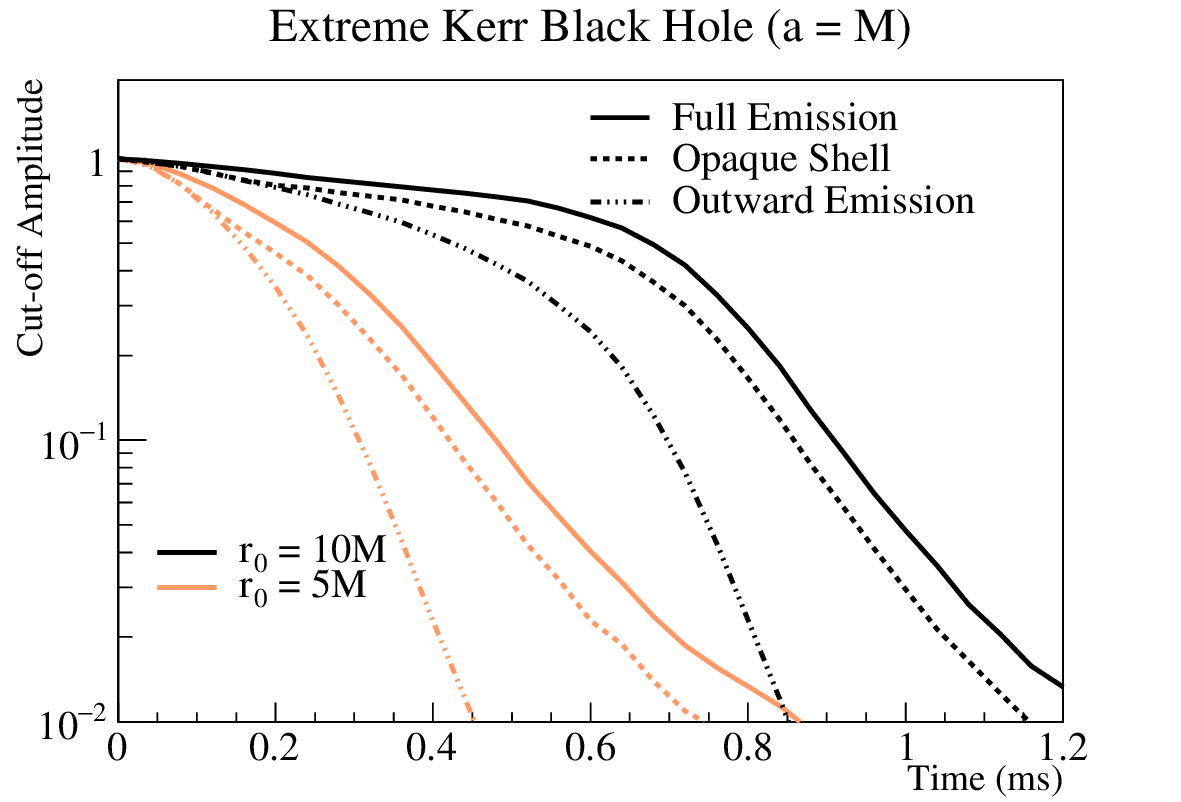}
    \caption{The three inward emission scenarios in an extreme Kerr black hole: transparent  medium,  with  emissions  allowed  in  all directions (solid); opaque inner medium, allowing only emissions which are outward in the $FF$ frame (dotted); and allowing only emissions which are outward in the LNRF, or $S$ frame (dashed-dotted).}
    \label{fig:m25_ker_opaque}
\end{figure}

The resulting cut-off profile is shown in Figs.~\ref{fig:cutoff_kerr_m25}
and \ref{fig:m25_ker_opaque}.
The long tail is not very evident in the sub-extremal case of $a=0.5M$,
but in the extremal case of $a=M$, the modification of the cut-off is more
significant.  It is also evident from
Fig.~\ref{fig:m25_ker_opaque} that the modification is mostly due to those
geodesics which are emitted inwards in the LNRF, but are outward in
the $FF$ frame.
Unlike in the non-rotating case, the decay parameter of the tail
does not approach a limiting value, but rather continues to increase in a
manner which may be noticeable even before the time the neutrinos are redshifted
below detectable energies.

\section{Summary and Discussion}
\label{sec:conc}

We have investigated the contribution of non-radial neutrino emissions to the
shape of the neutrino cut-off expected upon the formation of a black hole within
a core-collapse supernova.  Our toy calculations, based on ray-tracing null
geodesics from contracting matter shells or rings, show for the Schwarzschild case
the cut-off starting
with a slow decrease in luminosity over several tenths of milliseconds, depending
on the model of the shell and emitters' velocities, followed
by a rapid decrease which approaches an exponential decay with time constant
$3\sqrt{3}M$, a value calculated in the 1960s by
Podurets~\cite{podurets:64} and Ames and Thorne~\cite{AmesThorne:1968}.
The $3\sqrt{3}M$ time constant features in the luminosity profiles across different
modifications to the model, indicating that all such models end up with
neutrinos slowly leaking from near the radius $3M$.  If this part of the cut-off
can be resolved in time, it would represent an independent handle on the mass of the
newly formed black hole.

We estimate how many neutrino events may be available for resolving the time constant
by integrating a simple exponential tail, as if the cut-off begins abruptly rather
than turning over a ``knee'' as seen in Fig.~\ref{fig:m25_opaque} and elsewhere.
We use the $40\Msun$ model from Sec.~\ref{sec:vel}, observed at a distance of 10kpc.
Event rates for several neutrino detectors can be found in Fig.~3 of \cite{Gullin:2021hfv}.
Super-Kamiokande~\cite{superK:2003} and JUNO~\cite{juno:physics:16}
may be expected to see an event rate of around 15 per ms before the cut-off,
followed by a tail of approximately 0.7 events.  This yield is unlikely to
result in a measurement but is large enough that it introduces a systematic uncertainty
in how well the cut-off can be localized in time; this uncertainty would then feed
into applications such as using the observed cut-off in several detectors to
triangulate the direction to the CCSN.  On the other hand, the estimated event rate for
Hyper-Kamiokande~\cite{hyperk:dr:2018} before the cut-off is around 100 events per ms,
followed by a tail of 5 events,
which may indeed give a crude value even after taking
into account the uncertainty in when the exponential cut-off begins.
It should be noted, however, that these events likely will be
mixed with those of other effects, among them an ``echo'' of neutrinos scattering
off infalling material which is explored in a companion work~\cite{Gullin:2021hfv}.

In the Kerr case, we see that even for rotations as large as $a=0.5M$, the
non-radial geodesics appear to introduce a small delay to the cut-off, but do
not otherwise noticeably modify the tail or its time constant.  For extreme rotation, however,
the neutrino leakage from near the horizon extends the tail significantly,
and doubles the number of events expected to be observed.
Even though the extreme rotation case is usually considered to be unlikely,
it highlights the desirability of resolving the shape of the cutoff,
not just to measure rotation, but also to gauge the validity of using the
time constant as a mass measurement.

Even though this study is very simplistic, it suggests that if there is an abrupt
drop in the neutrino emission from a CCSN, signalling the formation of a black hole,
it is worth examining the shape of the cut-off in more detail from both
theoretical and observational perspectives.
The advent of the next generation's larger detectors, as well as the combination of
detectors via SNEWS, puts such measurements tantalizingly within reach.

The authors would like to thank Philipp Podsiadlowski, Shuai Zha, Ming-chung Chu, Luc Nguyen,
Steve Biller, Armin Reichold,
and the Oxford SNO+ group for stimulating and useful discussions.
This research is supported by the Science and Technology Facilities Council of
the United Kingdom (Grant No. ST/S000933/1)
and the Swedish Research Council (Project No. 2020-00452).

% from Samuel/Evan:
% (simulations refer to those in the other article)
%The simulations were enabled by resources provided by the Swedish
%National Infrastructure for Computing (SNIC) at PDC
%and NSC partially funded by the Swedish Research
%Council through grant agreement No. 2016-07213.

\newpage
\bibliography{references}% Produces the bibliography via BibTeX.

\end{document}